\pdfoutput=1

\documentclass[11pt, dvipsnames]{article}

\usepackage[preprint]{acl}

\usepackage{times}
\usepackage{latexsym}

\usepackage[T1]{fontenc}

\usepackage{amsmath}
\usepackage{adjustbox}
\usepackage{amssymb}
\usepackage{algorithmicx}
\usepackage{algpseudocode}

\AtBeginDocument{%
  }
\usepackage{graphicx}
\usepackage{subfig}

\usepackage[utf8]{inputenc}

\usepackage{microtype}

\usepackage{inconsolata}

\usepackage{algorithm}
\usepackage{algorithmicx}
\usepackage[table]{xcolor}

\usepackage{graphicx}
\usepackage{algorithm}
\usepackage{algpseudocode}
\usepackage{amsmath}
\usepackage{dblfloatfix} 
\usepackage{multirow}
\usepackage{booktabs}
\usepackage{verbatim}
\usepackage{balance}
\usepackage{array}
\usepackage{float}
\usepackage{adjustbox}
\usepackage{afterpage}
\usepackage[table]{xcolor}
\algdef{SE}[VARIABLES]{Variables}{EndVariables}
   {\algorithmicvariables}
   {\algorithmicend\ \algorithmicvariables}
\algnewcommand{\algorithmicvariables}{\textbf{Declare}}

\PassOptionsToPackage{table}{xcolor}

\usepackage{times}
\usepackage{latexsym}
\usepackage[T1]{fontenc}
\usepackage[utf8]{inputenc}
\usepackage{microtype}
\usepackage{inconsolata}
\usepackage{graphicx}
\usepackage[table]{xcolor}
\usepackage{amsmath}
\usepackage{amssymb}

\title{Knowledge Graph Enhanced Memory-Augmented Retrieval for Long Context Modeling}

\author{
  Ghadir Alselwi\textsuperscript{1},
  Basem Suleiman\textsuperscript{1}, 
  Hao Xue\textsuperscript{1,2},
  Shoaib Jameel\textsuperscript{3},
  Hakim Hacid\textsuperscript{4},\\
  \textbf{Flora D. Salim}\textsuperscript{1},
  \textbf{Imran Razzak}\textsuperscript{5,1} \\
  \textsuperscript{1}University of New South Wales, Sydney, NSW, Australia \\
  \textsuperscript{2}Hong Kong University of Science and Technology (Guangzhou) \\
  \textsuperscript{3}University of Southampton, Southampton, UK \\
  \textsuperscript{4}Technology Innovation Institute, Abu Dhabi, UAE \\
  \textsuperscript{5}Mohamed Bin Zayed University of Artificial Intelligence, Abu Dhabi, UAE \\
  \small \texttt{\{g.alselwi,b.suleiman,flora.salim\}@unsw.edu.au} \\
}

\begin{document}
\maketitle

\begin{abstract}
Long-context language modeling requires not only extending context windows but maintaining coherent understanding of entity states and relationships across thousands of tokens---a challenge that semantic similarity alone cannot address. KGERMAR addresses this by constructing dynamic, context-specific knowledge graphs from input text during inference, enabling domain-adaptive retrieval that leverages both semantic similarity and explicit entity relationships. The framework performs real-time entity and relation extraction to build contextual knowledge graphs, then integrates graph-structural embeddings with textual semantics through a multi-component memory architecture. Three memory banks---contextual, semantic, and structural---are maintained with retrieval signals fused via learned weights to capture both surface-level semantics and deeper relational patterns. Evaluated on SlimPajama (84.7K training examples), WikiText-103 (4,358 examples), PG-19 (100 examples), and Proof-pile (46.3K examples), KGERMAR achieves up to 8.5\% lower perplexity and 2--2.5x better memory efficiency than memory-augmented baselines across context lengths from 1K to 32K tokens, with superior in-context learning performance across five NLU tasks. The dynamic knowledge graph construction approach advances memory-augmented language modeling by enabling domain-specific knowledge representation that adapts to input contexts rather than relying on fixed knowledge bases.
\end{abstract}

\section{Introduction}
\label{introduction}

\begin{figure}
    \centering
     \includegraphics[width=0.88\columnwidth]{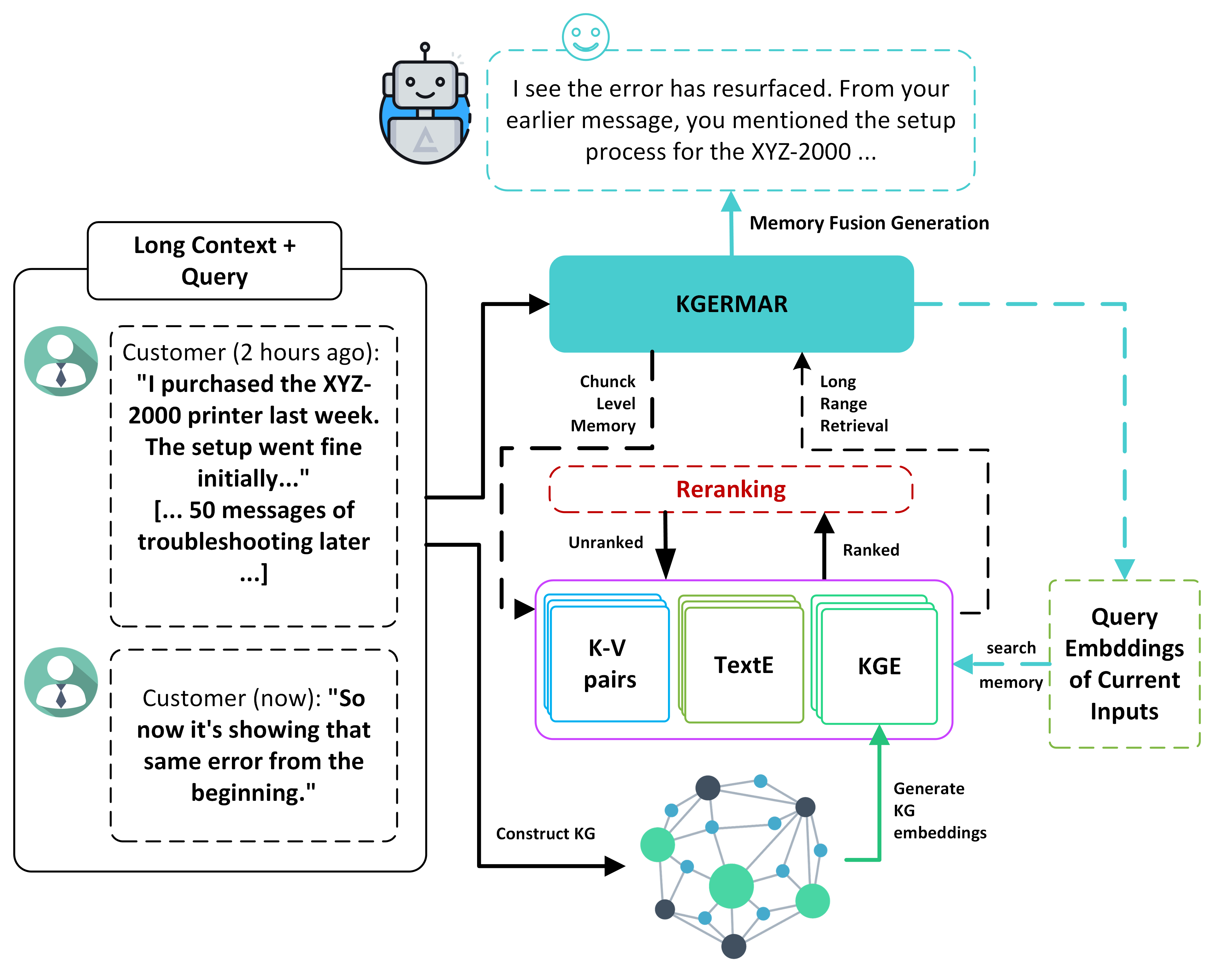}
    \caption{Overview of KGERMAR system showing dynamic knowledge graph construction from input text and multi-component memory architecture for retrieval.}
    \label{fig:toyexample}
\end{figure}

Long-context language modeling faces a challenge beyond simply extending context  windows: \textit{maintaining coherent understanding of entity states and relationships  across thousands of tokens}. Consider a technical support assistant processing a  50-message troubleshooting session, as illustrated in Figure~\ref{fig:toyexample}.  A customer reports at message 47: \textit{``So now it's showing that same error  from the beginning.''} To generate a useful response, the model must connect this  to message 1---\textit{``I purchased the XYZ-2000 printer last week. The setup went  fine initially''}---while tracking which solutions were attempted across the  intervening exchanges, which failed, and why. This is not merely a retrieval  precision problem; it is a \textit{structural reasoning problem}: the model must  reconstruct a causal chain of device states, actions, and outcomes distributed  across a context far exceeding what attention mechanisms reliably  handle~\cite{vaswani2017attention, tworkowski2024focused}.

Standard long-context approaches address this through attention window extension or  memory-augmented retrieval, but both share a critical blind spot. Extended-attention  models~\cite{chen2023longlora, peng2023yarn} suffer from well-documented  \textit{lost-in-the-middle} degradation~\cite{liu2024lost}: information in the  middle of a long context is systematically under-attended, regardless of its  relevance. Memory-augmented systems~\cite{borgeaud2022improving, liu2024memlong}  address attention scaling by retrieving relevant chunks, but rely exclusively on  \textit{semantic similarity}---returning passages that \textit{mention} the same  terms rather than passages that describe \textit{related states of the same entity}.  In the support example, a semantic retriever may surface passages mentioning  ``printer error'' without distinguishing whether those passages describe the  \textit{initial setup failure}, a \textit{subsequently resolved issue}, or an  \textit{unrelated error code}---precisely the distinctions needed for a coherent  response. These entity states and causal progressions are encoded not in token  co-occurrence patterns but in the \textit{structural relationships} between entities  across a document.

Current approaches to integrating structural knowledge with retrieval face three  critical limitations that make them unsuitable for this setting. First, most methods  rely on static, pre-constructed knowledge graphs like ConceptNet or Freebase that  cannot represent domain-specific or session-specific entities---the XYZ-2000 printer  model, its specific error codes, or the sequence of actions taken during this  particular support session~\cite{trivedi2017know, jin2019recurrent}. Second,  existing graph-neural retrieval approaches operate over these fixed general-purpose  graphs, providing no mechanism to extract and reason over relationships that only  emerge from the content at hand~\cite{xiong2017deeppath, das2017go}. Third,  integrating on-the-fly knowledge graph construction with neural retrieval for  long-context inference remains largely unexplored~\cite{sun2019pullnet,  xiao2019dynamically}.

We present \textbf{KGERMAR}, a framework that directly addresses these failures  by constructing \textit{dynamic, context-specific knowledge graphs} during inference  and integrating graph-structural embeddings with textual semantics for improved  long-context modeling. The key insight is that the same input text used for language  modeling contains rich structural information---entities, their types, and the  relationships between them---that can be extracted in real-time to build a  domain-adaptive knowledge graph. Unlike static-graph approaches~\cite{trivedi2017know,  xiong2017deeppath}, KGERMAR captures relationships specific to each document  collection as it processes them. Unlike purely semantic systems including  ERMAR~\cite{alselwi2025long} and MemLong~\cite{liu2024memlong}, KGERMAR maintains  three specialized memory components---contextual, semantic, and structural---enabling  retrieval that distinguishes between passages mentioning related terms coincidentally  versus passages describing explicit relationships between the same entities. This  distinction is critical precisely in long-context settings, where the relevant  passage may appear thousands of tokens earlier and share only structural, not lexical,  similarity with the query.

\begin{figure*}
    \centering
    \includegraphics[scale=0.26]{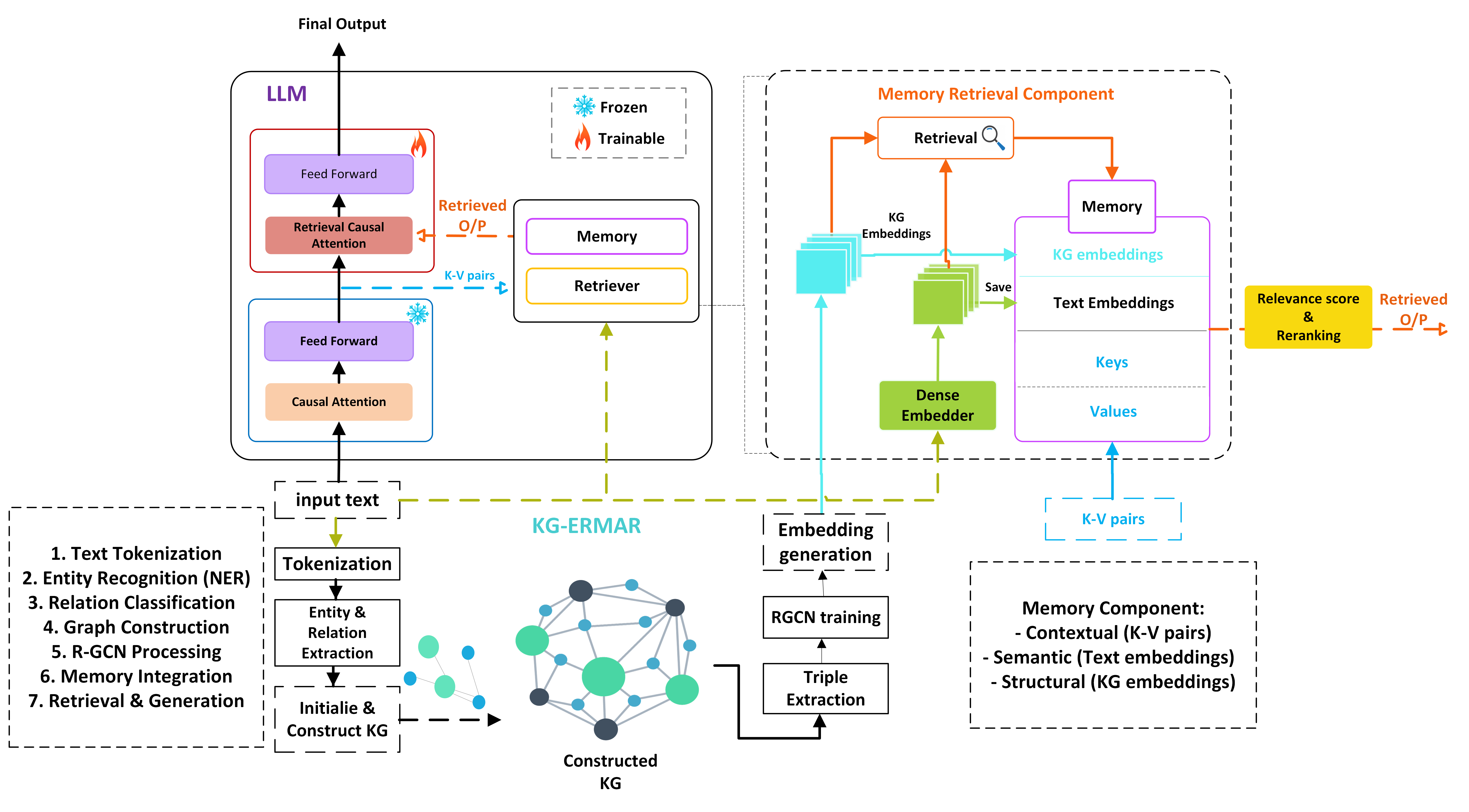}
    \caption{Architecture diagram illustrating the four main components: contextual knowledge graph construction, graph-enhanced embedding generation, multi-component memory architecture, and multi-modal retrieval mechanism.}
    \label{fig:KGPart}
\end{figure*}

As illustrated in Figures~\ref{fig:toyexample} and~\ref{fig:KGPart}, KGERMAR  processes input text through an integrated pipeline: (1) transformer-based NER and  relation extraction identify entities and relationships, constructing a contextual  knowledge graph; (2) Relational Graph Convolutional Networks  (R-GCN)~\cite{schlichtkrull2018modeling} encode graph structure into entity  embeddings capturing both individual properties and structural roles; (3) cross-modal  attention fuses graph-structural embeddings with textual entity embeddings from the  base language model; (4) three memory banks store contextual key-value pairs  (following ERMAR), semantic text embeddings, and structural entity embeddings,  with retrieval signals combined via learned fusion weights. Crucially, the structural  memory bank enables the model to retrieve the passage describing the \textit{initial  printer setup} in response to a query about a \textit{recurring error}---not because  the two passages share vocabulary, but because both involve the same device entity  in a causally connected state progression.

\textbf{Contributions.} This work advances memory-augmented retrieval for 
long-context modeling through three key innovations:
\textbf{(i)} A dynamic knowledge graph construction pipeline that extracts 
domain-specific entities and relationships from input text in real-time during inference, enabling adaptation to specialized and session-specific content absent 
from static knowledge bases;
\textbf{(ii)} A hybrid multi-component memory architecture integrating contextual, semantic, and structural representations, where graph-structural embeddings capture entity relationships that are complementary to---and often more discriminative than---the semantic similarity signals used by existing memory-augmented systems;
\textbf{(iii)} Empirical demonstration that incorporating structural knowledge yields substantial improvements---up to 8.5\% lower perplexity and 2--2.5$\times$ better memory efficiency across context lengths from 1K to 32K tokens---despite increased inference latency from real-time graph construction, for which we provide detailed analysis to inform deployment tradeoffs.

Extensive experiments on SlimPajama (84.7K training examples) for fine-tuning, and  WikiText-103 (4,358 examples), PG-19 (100 examples), and Proof-Pile (46.3K examples)  for evaluation demonstrate that KGERMAR consistently outperforms strong baselines  including ERMAR~\cite{alselwi2025long} and MemLong~\cite{liu2024memlong}, while  also showing competitive memory efficiency against extended-context models.

\section{KG-Enhanced Memory-Augmented Retrieval}
\label{sec:method}

KGERMAR enhances memory-augmented retrieval by constructing dynamic, context-specific knowledge graphs and integrating graph-structural embeddings with textual semantics. Unlike approaches relying on static knowledge bases~\cite{xiong2017deeppath, das2017go} or purely semantic  similarity~\cite{alselwi2025long, liu2024memlong}, KGERMAR extracts entities  and relationships directly from input text, leveraging both textual representations and graph structure for enhanced long-context retrieval.

\subsection{Framework Overview}
\label{subsec:overview}

KGERMAR comprises four integrated components: (1)~\textbf{Contextual Knowledge  Graph Construction} (\S\ref{subsec:kg_construction}) performs real-time entity  and relation extraction to build domain-specific knowledge structures;  (2)~\textbf{Graph-Enhanced Embedding Generation} (\S\ref{subsec:embeddings})  employs R-GCN~\cite{schlichtkrull2018modeling} to encode graph structure, then  fuses these with textual embeddings via cross-modal attention;  (3)~\textbf{Multi-Component Memory Architecture} (\S\ref{subsec:memory}) maintains  three specialized memory banks---contextual (key-value pairs from language model  layers), semantic (dense text embeddings), and structural (graph-enhanced entity  embeddings)---each updated incrementally as new contexts arrive and managed via  LRU eviction when capacity is reached; (4)~\textbf{Hybrid Retrieval Mechanism}  (\S\ref{subsec:memory}) combines retrieval signals from all three banks via learned  fusion weights, with retrieved representations injected into the LM through  retrieval causal attention at designated upper layers, following the  cross-attention injection design of ERMAR~\cite{alselwi2025long}. The key distinction from ERMAR~\cite{alselwi2025long} is the addition of structural  memory and dynamic knowledge graph construction, enabling the system to distinguish  between passages mentioning related terms versus passages describing explicit entity  relationships. Figure~\ref{fig:KGPart} illustrates the complete pipeline.

\subsection{Contextual Knowledge Graph Construction}
\label{subsec:kg_construction}

Given input contexts $\mathcal{D} = \{d_1, d_2, \ldots, d_n\}$ where  $d_i = (w_{i,1}, w_{i,2}, \ldots, w_{i,|d_i|})$ contains tokens $w_{i,k}$ from  vocabulary $\mathcal{V}$, we construct a contextual knowledge graph  $\mathcal{G}_i = (\mathcal{E}_i, \mathcal{R}, \mathcal{T}_i)$ for each context,  where $\mathcal{E}_i$ represents extracted entities, $\mathcal{R}$ denotes relation  types, and $\mathcal{T}_i \subseteq \mathcal{E}_i \times \mathcal{R} \times  \mathcal{E}_i$ contains relation triples.

\textbf{Entity Recognition.} We use BERT-large fine-tuned on CoNLL-2003 as the NER model, identifying entity spans, types, and positions:
\begin{equation}
\small
\begin{aligned}
\mathcal{E}_i = \{
&(e_j, \tau_j, s_j, t_j) :
e_j \in \mathcal{E},\;
\tau_j \in \mathcal{T}_{entity}, \\
&s_j \le t_j \in [1, |d_i|]
\}
\end{aligned}
\end{equation}
where $e_j$ is the entity text, {\small $\tau_j \in \mathcal{T}_{entity} =  \{\textsc{Person}, \textsc{Organization}, \textsc{Location}, \textsc{Concept},  \textsc{Event}\}$} is the entity type, and $(s_j, t_j)$ denote start and end  token positions within context $d_i$. These five types provide broad coverage  across the technical, literary, and scientific domains of our evaluation datasets.

\textbf{Relation Extraction.} We use BERT-base fine-tuned on TACRED~\cite{zhang2017position} 
as the relation extraction model $\mathcal{F}_r$. For entity pairs $(e_h, e_t)$ within a context window of $W$ tokens, $\mathcal{F}_r$ predicts:
\begin{equation}
r_{h,t} = \mathcal{F}_r(e_h, e_t, \text{context}_{W}(e_h, e_t, d_i))
\end{equation}
where $\text{context}_{W}(e_h, e_t, d_i)$ extracts a $W$-token window around the  entity pair from document $d_i$, producing triples $(e_h, r_{h,t}, e_t)$ where  $r_{h,t} \in \mathcal{R} \cup \{\emptyset\}$ ($\emptyset$ indicates no relation). 
The supported relation types are \textsc{HasProperty}, \textsc{CausedBy}, \textsc{PartOf}, \textsc{UsedFor}, \textsc{IsA}, \textsc{LocatedAt}, and \textsc{RelatedTo}, covering causal, compositional, and functional relationships salient in long-form technical and scientific text. The fixed window $W$ limits extraction to local entity pairs; cross-paragraph  relations are approximated through graph propagation in R-GCN layers  (\S\ref{subsec:embeddings}), which we discuss further in the Limitations section.

\textbf{Graph Construction.} Three steps improve graph quality, motivated by standard practices in neural information extraction~\cite{li2019entity, zheng2017joint}: (1)~\textit{Entity consolidation} merges variant mentions of the same entity using both string similarity (Levenshtein distance) and embedding similarity (cosine similarity between base LM representations):
\begin{equation}
\small
\begin{aligned}
\text{Merge}(e_i, e_j) \iff \;&\text{sim}_{string}(e_i, e_j) > \theta_{str} \\
&\lor\; \text{sim}_{embed}(e_i, e_j) > \theta_{ctx}
\end{aligned}
\end{equation}
with thresholds $\theta_{str} = 0.85$ and $\theta_{ctx} = 0.90$ selected via grid search on the SlimPajama validation set; (2)~\textit{Relation confidence scoring} combines classifier certainty and within-context frequency:
\begin{equation}
\small
\begin{split}
w(h, r, t) = &\, \alpha \cdot \text{conf}_{\mathcal{F}_r}(h, r, t) \\
&+ (1-\alpha) \cdot \frac{\text{freq}(h, r, t)}{\max_{(h',r',t')} 
\text{freq}(h', r', t')}
\end{split}
\end{equation}
where $\text{conf}_{\mathcal{F}_r}(h, r, t)$ is the softmax probability from  $\mathcal{F}_r$, $\text{freq}(h, r, t)$ counts repeated occurrences of the  triple pattern within the current context, and $\alpha = 0.7$ weights classifier  confidence more heavily than frequency; (3)~\textit{Graph filtering} removes  low-confidence triples: $\mathcal{T}_i = \{(h, r, t) : w(h, r, t) > \theta_{conf}\}$  where $\theta_{conf} = 0.5$ balances precision and recall, also determined on  the validation set. Algorithm~\ref{alg:kg_construction}  (Appendix~\ref{sec:appendix}) details the complete procedure.

\subsection{Graph-Enhanced Embedding Generation}
\label{subsec:embeddings}

\textbf{R-GCN Encoding.} For entity $e_i$, the embedding $\mathbf{h}_i^{(l+1)} \in \mathbb{R}^{d^{(l+1)}}$ at layer $l+1$ is computed via relation-specific message passing:
\small
\begin{equation}
\small
\mathbf{h}_i^{(l+1)} = \sigma \left( \mathbf{W}_0^{(l)} \mathbf{h}_i^{(l)} + 
\sum_{r \in \mathcal{R}} \sum_{j \in \mathcal{N}_i^r} \frac{1}{|\mathcal{N}_i^r|} 
\mathbf{W}_r^{(l)} \mathbf{h}_j^{(l)} \right)
\end{equation}
where {\small $\mathcal{N}_i^r = \{j : (j, r, i) \in \mathcal{T}_i \lor (i, r, j)  \in \mathcal{T}_i\}$} denotes neighbors of entity $i$ via relation $r$ (both  incoming and outgoing edges), {\small $\mathbf{W}_0^{(l)} \in \mathbb{R}^{d^{(l+1)}  \times d^{(l)}}$} is the self-connection weight matrix, {\small $\mathbf{W}_r^{(l)}  \in \mathbb{R}^{d^{(l+1)} \times d^{(l)}}$} is the relation-specific weight  matrix, and $\sigma$ is the ReLU activation. Multi-hop propagation across R-GCN  layers allows entities to aggregate information beyond the local extraction  window $W$, partially compensating for the window constraint in relation  extraction. This enables entities to aggregate neighborhood information,  reflecting both their properties and relationship types.

\textbf{Hybrid Text-Graph Embeddings.} Each entity is encoded using the base  language model to capture contextual semantics:  {\small $\mathbf{e}_{text}^{(i)} = \text{MeanPool}(\text{LM}(\text{context}_{C}(e_i, d)))$}  where $\text{context}_{C}(e_i, d)$ extracts $C = 64$ tokens surrounding entity  $e_i$ in document $d$, $\text{LM}(\cdot)$ applies OpenLLaMA-3B, and  $\text{MeanPool}(\cdot)$ averages token representations over the entity span.  Cross-modal fusion combines both modalities:
\small
\begin{align}
\mathbf{a}_{i} &= \text{softmax}\left(\frac{\mathbf{h}_i^{(L)} \cdot 
\mathbf{e}_{text}^{(i)}}{\sqrt{d_{kg}}}\right) \\
\mathbf{e}_i &= \mathbf{W}_{fusion} \begin{bmatrix} \mathbf{h}_i^{(L)} \\ 
\mathbf{e}_{text}^{(i)} \end{bmatrix} + \mathbf{a}_{i} \cdot 
(\mathbf{h}_i^{(L)} + \mathbf{e}_{text}^{(i)})
\end{align}
where $\mathbf{h}_i^{(L)}$ is the graph-structural embedding from the final  R-GCN layer, $\mathbf{e}_{text}^{(i)}$ is the textual embedding, $\mathbf{a}_i$  is an attention weight measuring compatibility between the two modalities,  $\mathbf{W}_{fusion} \in \mathbb{R}^{d_{final} \times (d_{kg} + d_{text})}$ is  a learned projection matrix, and $\mathbf{e}_i$ is the final fused entity  representation stored in structural memory.

\subsection{Multi-Modal Memory Architecture and Retrieval}
\label{subsec:memory}

\textbf{Memory Components.} KGERMAR maintains three specialized banks with  distinct granularity, content, and update policies. \textit{Contextual memory}  $\mathcal{M}_{ctx} = \{(\mathbf{k}_i, \mathbf{v}_i)\}$ stores token-level  key-value pairs extracted from layer 13 of the 26-layer LM following  ERMAR~\cite{alselwi2025long}, updated at every forward pass with LRU eviction  at capacity 32,768. \textit{Semantic memory} $\mathcal{M}_{sem} = \{(\mathbf{s}_i, c_i)\}$  stores BGE-M3~\cite{borgeaud2022improving} embeddings of 256-token non-overlapping  chunks, updated per chunk with FAISS HNSW indexing for efficient retrieval.  \textit{Structural memory} $\mathcal{M}_{kg} = \{(\mathbf{e}_i, \text{entity}_i)\}$  stores the fused text-graph embeddings from Equation~(8), updated per context  window as the knowledge graph is constructed. Retrieved representations from  all three banks are injected into the LM via retrieval causal attention at  layers $\{14, 18, 22, 26\}$, following the cross-attention mechanism  of ERMAR~\cite{alselwi2025long}.

\textbf{Hybrid Retrieval.} For query $q$, parallel top-$K$ retrieval is performed from each bank:
\small
\begin{align}
\mathcal{R}_{ctx}(q) &= \text{TopK}\left(\text{sim}_{dot}
(\text{Embed}(q), \mathbf{K}_{ctx})\right) \\
\mathcal{R}_{sem}(q) &= \text{TopK}_{\text{FAISS}}\left(\text{sim}_{cos}
(\text{Embed}(q), \mathbf{S})\right) \\
\mathcal{R}_{kg}(q) &= \text{TopK}_{\text{FAISS}}\left(\text{sim}_{cos}
(\text{Embed}(q), \mathbf{E}_{kg})\right)
\end{align}
where $\mathbf{K}_{ctx}$ are contextual keys, $\mathbf{S}$ are semantic  embeddings, $\mathbf{E}_{kg}$ are knowledge graph embeddings, and  $\text{Embed}(q)$ is the query embedding. Contextual memory uses dot product  similarity matching ERMAR, while semantic and structural memory use cosine  similarity. FAISS HNSW indexing~\cite{borgeaud2022improving} enables  $O(\log |\mathcal{M}|)$ query complexity. Final scoring combines all signals:
\begin{equation}
\small
\begin{aligned}
\text{score}(q, r_i) =
\lambda_1 \cdot \text{sim}_{ctx}(q, r_i)
+ \lambda_2 \cdot \text{sim}_{sem}(q, r_i) \\
+ \lambda_3 \cdot \text{sim}_{kg}(q, r_i)
\end{aligned}
\end{equation}
where fusion weights $\{\lambda_1, \lambda_2, \lambda_3\}$ are learned via gradient descent with $\lambda_1 + \lambda_2 + \lambda_3 = 1$. Algorithm~\ref{alg:multimodal_retrieval} (Appendix~\ref{sec:appendix}) details the complete procedure.

\subsection{Training Objectives}
\label{subsec:training}

KGERMAR employs multi-task learning optimizing four objectives:
\begin{equation}
\small
\mathcal{L}_{total} = \mathcal{L}_{LM} + \alpha \cdot \mathcal{L}_{KG} + 
\beta \cdot \mathcal{L}_{retrieval} + \gamma \cdot \mathcal{L}_{alignment}
\end{equation}
where $\alpha = 0.1$, $\beta = 0.05$, $\gamma = 0.01$ control the relative 
importance of each auxiliary objective relative to the primary language 
modeling loss.
\textit{Language modeling loss} optimizes next-token prediction with retrieved context: $\mathcal{L}_{LM} = -\frac{1}{|D|}\sum_{i=1}^{|D|} \log p_\theta (x_i | x_{<i}, \mathcal{R}(x_{<i}))$ where $\mathcal{R}(x_{<i})$ is the retrieved context for prefix $x_{<i}$.

\textit{Knowledge graph embedding loss} trains R-GCN via link prediction with binary cross-entropy:
\small
\begin{align}
\mathcal{L}_{KG} = &\sum_{(h,r,t) \in \mathcal{T}} 
\text{BCE}(\sigma(\mathbf{h}_h^{(L)\top} \mathbf{W}_r \mathbf{h}_t^{(L)}), 1) \nonumber \\
&+ \sum_{(h,r,t') \in \mathcal{T}_{neg}} 
\text{BCE}(\sigma(\mathbf{h}_h^{(L)\top} \mathbf{W}_r \mathbf{h}_{t'}^{(L)}), 0)
\end{align}
where $\mathcal{T}$ contains positive triples, $\mathcal{T}_{neg}$ contains negative samples with randomly replaced tail entities, and $\sigma$ is sigmoid.

\textit{Retrieval ranking loss} uses margin-based ranking:  $\mathcal{L}_{retrieval} = \sum_{q,r^+,r^-} \max(0, \epsilon +  \text{score}(q,r^-) - \text{score}(q,r^+))$ where $r^+$ and $r^-$ are  relevant and irrelevant retrieved contexts and $\epsilon = 0.1$ is the margin.

\textit{Cross-modal alignment loss} ensures consistency between textual and structural representations: $\mathcal{L}_{alignment} = \frac{1}{|\mathcal{E}|} \sum_{e \in \mathcal{E}} ||\mathbf{h}_e^{(L)} - \mathbf{e}_{text}^{(e)}||_2^2$.

\section{Related Work}
\label{sec:related_work}

We position KGERMAR within three research areas: memory-augmented retrieval for long contexts, knowledge graph methods, and knowledge-enhanced language models.

\subsection{Memory-Augmented Retrieval for Long Contexts}

Dense retrieval methods~\cite{karpukhin2020dense, khattab2020colbert} learn  semantic similarity in continuous embedding spaces. RAG~\cite{lewis2020retrieval}  pioneered integrating dense retrieval with generative models, enabling external  memory access beyond model parameters. For long-context modeling,  RETRO~\cite{borgeaud2022improving} demonstrated retrieval from large corpora,  MemLong~\cite{liu2024memlong} introduced chunk-level memory with side networks,  and ERMAR~\cite{alselwi2025long} advanced this with ranking mechanisms over  intermediate LM key-value pairs. Agentic systems like Mem0~\cite{chhikara2025mem0}  and A-Mem~\cite{xu2025mem} extend RAG for interactive agents but focus on  cross-session persistence rather than within-document structural reasoning.  Extended-attention models~\cite{chen2023longlora, peng2023yarn} and  long-context LLMs~\cite{abdin2024phi} make more tokens accessible but do not  improve retrieval \textit{precision}---they cannot prioritize structurally relevant  passages over semantically coincident ones, nor do they offer the memory efficiency  gains (2--2.5$\times$ lower peak GPU memory at 32K tokens) that retrieval-based  architectures provide on commodity hardware. Critically, all memory-augmented  systems rely exclusively on semantic similarity without modeling entity  relationships~\cite{ji2021survey, wang2021kepler, yasunaga2021qa}, limiting  precision in knowledge-intensive scenarios where entity states and causal  relationships determine relevance.

\subsection{Knowledge Graphs and Graph-Enhanced Models}
Knowledge graph construction has evolved from rule-based systems like  NELL~\cite{mitchell2018never} and OpenIE~\cite{etzioni2008open} to neural  approaches~\cite{li2019entity, zheng2017joint}, while graph representation  methods including TransE~\cite{bordes2013translating}, RotatE~\cite{sun2019rotate},  and R-GCN~\cite{schlichtkrull2018modeling} encode pre-constructed graphs into  dense embeddings. Knowledge-enhanced language models---ERNIE~\cite{sun2019ernie},  KnowBERT~\cite{peters2019knowledge}, K-BERT~\cite{liu2020k},  GreaseLM~\cite{zhang2022greaselm}, and GraphFormers~\cite{yang2021graphformers}---integrate  structured knowledge during pre-training. Graph-enhanced retrieval systems  QA-GNN~\cite{yasunaga2021qa} and DRAGON~\cite{yasunaga2022deep} apply GNNs  over ConceptNet for question answering, while earlier work used static knowledge  bases for query expansion~\cite{dalton2014entity, xiong2017explicit}. All these  approaches share a critical constraint: they rely on static, general-purpose  knowledge graphs with fixed vocabularies, making them ineffective when relevant  entities and relationships must be derived from specialized content  itself~\cite{trivedi2017know, jin2019recurrent, sun2019pullnet, xiao2019dynamically}.

\subsection{Positioning of KGERMAR}
KGERMAR occupies a distinct position at the intersection of memory-augmented  retrieval and dynamic knowledge graph construction. Unlike static graph  approaches~\cite{yasunaga2021qa, yasunaga2022deep, sun2019ernie}, it constructs  domain-specific knowledge graphs from input text during inference. Unlike  memory-augmented systems~\cite{liu2024memlong, alselwi2025long}, it captures  explicit entity relationships alongside semantic similarity. Unlike  knowledge-enhanced models~\cite{peters2019knowledge, zhang2022greaselm}, its  structural knowledge adapts to each document at inference time. Building on  ERMAR~\cite{alselwi2025long}, KGERMAR augments contextual memory with semantic  memory (BGE-M3 embeddings~\cite{borgeaud2022improving}) and structural memory  (graph-enhanced entity embeddings), producing a retrieval signal space strictly  richer than any prior approach---applicable across technical support, scientific  literature, and legal document analysis~\cite{ji2021survey, petroni2019language,  sun2019ernie}.

\section{Experimental Setup}

\subsection{Datasets}
We fine-tuned KGERMAR on \textit{SlimPajama}~\cite{fu2024data}, a high-quality  deduplicated corpus spanning web, book, and code domains, designed for long-context  training. It contains 84.7K training examples, preprocessed with a sliding window  of 512-token strides to ensure coverage of long sequences.

Evaluation was conducted on three benchmark datasets representing distinct  long-context domains: \textit{WikiText-103}~\cite{merity2016pointer} (4,358 test  examples), a structured encyclopedia corpus that benefits from entity-relationship  modeling; \textit{PG-19}~\cite{rae2019compressive} (100 test examples), a  book-length narrative corpus requiring long-range dependency tracking; and  \textit{Proof-Pile}~\cite{azerbayev2023proofpile} (46.3K test examples), a  mathematical and scientific text corpus testing performance on formal,  knowledge-dense content. These three datasets were selected to evaluate KGERMAR  across structurally diverse long-context settings, following the evaluation  protocol of MemLong~\cite{liu2024memlong} and ERMAR~\cite{alselwi2025long}.  Perplexity is computed on the final 2048 tokens of sequences ranging from 1024  to 32768 tokens~\cite{yen2024long}.

\subsection{Model Configuration}

We fine-tuned OpenLLaMA-3B~\cite{touvron2023llama} using LoRA~\cite{hu2021lora}  for parameter-efficient training. The model comprises $L=26$ transformer layers  and $H=32$ attention heads with rotational position encoding~\cite{su2024roformer}.  Layer 13 serves as the memory layer storing historical key-value pairs, while  layers $\{14, 18, 22, 26\}$ are augmented with retrieval causal attention.  KGERMAR employs a memory capacity of 32,768 key-value pairs with BGE-M3  embeddings~\cite{borgeaud2022improving} for semantic similarity computation.  For knowledge graph construction, named entity recognition uses BERT-large  fine-tuned on CoNLL-2003, and relation extraction uses BERT-base fine-tuned  on TACRED~\cite{zhang2017position}, consistent with the model specification  in Section~\ref{sec:method}.

\subsection{Baseline Models}
KGERMAR was evaluated against state-of-the-art models at two parameter scales.  At 7B parameters: LLaMA-2-7B~\cite{touvron2023llama2} as a standard transformer  baseline; LongLoRA-7B-32k~\cite{chen2023longlora} using sparse attention for  32K-token contexts; and YARN-128k-7B~\cite{peng2023yarn} with dynamic position  embeddings supporting up to 128K tokens. At 3B parameters:  OpenLLaMA-3B~\cite{touvron2023llama} as the base architecture;  LongLLaMA-3B~\cite{tworkowski2024focused} in two retrieval configurations  (4 and 18 memory entries); MemLong-3B~\cite{liu2024memlong} as our primary  memory-augmented baseline; ERMAR~\cite{alselwi2025long} as our direct predecessor;  and Phi3-128k~\cite{abdin2024phi}, a strong non-memory-augmented long-context  model included to contextualize the tradeoffs between memory-augmented retrieval  and extended-attention approaches. This suite covers sparse attention, position  encoding extension, and memory-augmented strategies. Phi3-128k is included in  perplexity comparisons; its omission from ICL experiments follows the original  ERMAR evaluation protocol~\cite{alselwi2025long} and reflects the absence of  published ICL results for this model under comparable settings.

\subsection{Evaluation Metrics}

Perplexity on the final 2048 tokens serves as the primary metric for long-context  language modeling, focusing evaluation on long-range dependency modeling rather  than local context. For in-context learning, we report accuracy on five NLU  benchmarks---SST-2, MR, Subj, SST-5, and MPQA---in both 4-shot and 20-shot  settings, following MemLong~\cite{liu2024memlong}. Memory efficiency is assessed  via peak and reserved GPU memory usage and normalized memory per token.  Computational overhead is measured via per-token inference latency and throughput  (tokens per second) across context lengths from 1K to 32K tokens.

\section{Results and Discussion} 
\label{results}

\subsection{Long-Context Language Modeling}
Table~\ref{tab:LCLM} presents mean perplexity scores across sequence lengths  and datasets, following the evaluation protocol of~\cite{liu2024memlong}.  Lower perplexity indicates stronger language modeling capability.

Compared to fully fine-tuned models, OpenLLaMA-3B and LLaMA-2-7B demonstrate  competitive performance when test lengths remain within their pre-trained limits  (2048 and 4096 tokens respectively). However, once test lengths exceed these  limits, both models fail to generalize and exhibit significantly increased memory  overhead due to the quadratic complexity of attention. In contrast, KGERMAR  continues to reduce perplexity beyond its fine-tuning length of 1024 and  pre-trained length of 2048, demonstrating superior generalizability through  its retrieval-based memory architecture.

Among 7B models, YARN-128k-7B delivers stronger performance at shorter contexts  (1K--4K tokens) while LongLoRA-7B-32k maintains better stability at 16K.  Although LongLoRA's Shifted Sparse Attention significantly reduces memory usage,  it also diminishes performance on shorter texts. LongLLaMA, which can also store  K-V pairs, suffers from OOM issues at excessive lengths due to its infinitely  growing memory. Positional encoding models like YARN show strong generalization  capabilities, but can only guarantee that generation performance over long  distances does not degrade---they do not improve retrieval precision or control  memory consumption. Compared to these methods, KGERMAR leverages an external  multi-component memory architecture to handle longer input tokens while achieving  better perplexity improvements and effectively controlling GPU usage to avoid OOM  problems.

Among 3B models, Phi3-128k achieves the lowest perplexity on WikiText-103  (1K/2K/4K) and Proof-pile (4K), demonstrating that extended-attention training  is highly competitive on raw perplexity. However, Phi3-128k requires a dual-GPU  setup at 16K tokens and incurs 2--2.5$\times$ higher peak memory than KGERMAR  at equivalent context lengths (Table~\ref{tab:memory_efficiency}), highlighting  the practical deployment advantage of memory-augmented retrieval on commodity  hardware.

KGERMAR achieves the lowest perplexity among all memory-augmented 3B models  across configurations, outperforming both MemLong and ERMAR on WikiText-103  at all context lengths (7.74/7.25/7.20/7.14 vs.\ ERMAR's 8.42/7.61/7.62/7.80).  The WikiText-103 gains are consistent with the structured, entity-rich nature  of encyclopedia text, where knowledge graph construction captures cross-sentence  entity relationships that semantic similarity alone misses. On Proof-pile,  KGERMAR is competitive with MemLong at 2K and 4K but does not lead at 1K,  likely because mathematical text contains fewer named entities amenable to  NER-based graph construction, reducing the contribution of the structural memory  bank. PG-19  performance remains stable across all context lengths, demonstrating KGERMAR's  scalability on book-length narrative text requiring long-range dependency tracking.

\begin{table*}[htbp]
\centering
\scriptsize
\begin{tabular}{p{1.99cm}|p{.5cm}p{.5cm}p{.6cm}p{.59cm}|p{.5cm}p{.5cm}p{.6cm}p{.59cm}|p{.5cm}p{.5cm}p{.6cm}p{.59cm}}
\hline
& \multicolumn{4}{c|}{\textbf{PG19}} & \multicolumn{4}{c|}{\textbf{Proof-pile}} & \multicolumn{4}{c}{\textbf{WikiText-103}} \\
Model & 1k & 2k & 4k & 16k & 1k & 2k & 4k & 16k & 1k & 2k & 4k & 16k \\
\hline
\multicolumn{13}{c}{\cellcolor{gray!25}7B Model} \\
\hline
YARN-128k-7B & \textbf{7.22} & \textbf{7.47} & \textbf{7.17} & - & \textbf{3.03} & \textbf{3.29} & 2.98 & - & \textbf{5.71} & \textbf{6.11} & 5.71 & - \\
LongLoRA-7B-32k & 9.76 & 9.71 & 10.37 & \textbf{7.62} & 3.68 & 3.35 & \textbf{3.23} & \textbf{2.60} & 7.99 & 7.83 & 8.39 & \textbf{5.47} \\
LLaMA-2-7B & 10.82 & 10.06 & 8.92 & - & 3.24 & 3.40 & 2.72 & - & 10.82 & 6.49 & \textbf{5.66} & - \\
\hline
\multicolumn{13}{c}{\cellcolor{gray!25}3B Model} \\
\hline
Phi3-128k & 11.31 & 9.90 & \textbf{9.66} & -/9.65 & 4.25 & 3.11 & \textbf{2.77} & -/3.08 & \textbf{7.54} & \textbf{7.22} & \textbf{7.01} & -/7.20 \\
OpenLLaMA-3B & 11.60 & 9.77 & \(\scalebox{0.8}{$>$}\)\(\scalebox{0.9}{$10^3$}\) & - & \textbf{2.96} & \textbf{2.70} & \(\scalebox{0.8}{$>$}\)\(\scalebox{0.9}{$10^3$}\) & - & 10.57 & 8.08 & \(\scalebox{0.8}{$>$}\)\(\scalebox{0.9}{$10^3$}\) & - \\
LongLLaMA-3B* & 10.59 & 10.02 & \(\scalebox{0.8}{$>$}\)\(\scalebox{0.9}{$10^3$}\) & - & 3.55 & 3.15 & \(\scalebox{0.8}{$>$}\)\(\scalebox{0.9}{$10^3$}\) & - & 8.88 & 8.07 & \(\scalebox{0.8}{$>$}\)\(\scalebox{0.9}{$10^3$}\) & - \\
LongLLaMA-3B$^\dagger$ & 10.59 & 10.25 & 9.87 & - & 3.55 & 3.22 & 2.94 & - & 10.69 & 8.33 & 7.84 & - \\
MemLong-3B* & 10.66 & 10.09 & \(\scalebox{0.8}{$>$}\)\(\scalebox{0.9}{$10^3$}\) & - & 3.58 & 3.18 & \(\scalebox{0.8}{$>$}\)\(\scalebox{0.9}{$10^3$}\) & - & 8.72 & 7.93 & \(\scalebox{0.8}{$>$}\)\(\scalebox{0.9}{$10^3$}\) & - \\
w/4K MemLong & 10.54 & 9.95 & 9.89 & \textbf{9.64} & 3.53 & 3.16 & 3.15 & \textbf{2.99} & 8.53 & 7.92 & 7.87 & 7.99 \\
w/4K ERMAR & 10.32 & 9.75 & 9.78 & 9.81 & 3.24 & 2.98 & 3.03 & 3.18 & 8.42 & 7.61 & 7.62 & 7.80 \\
w/4K \textbf{KGERMAR} & \textbf{10.24} & \textbf{9.68} & 9.72 & 9.75 & 3.55 & 3.15 & 3.12 & 3.17 & 7.74 & 7.25 & 7.20 & \textbf{7.14} \\
\hline
\end{tabular}
\caption{Perplexity comparison across PG-19, Proof-pile, and WikiText-103 using sliding window evaluation. \textbf{Bold} indicates the lowest perplexity within each model group (7B and 3B) separately. Among 3B memory-augmented models, bold indicates the lowest perplexity; Phi3-128k achieves lower perplexity in several configurations but requires a dual-GPU setup at 16K tokens. ``-'' denotes OOM; ``x/y'' indicates single/dual GPU results. $*$~no extended memory; $\dagger$~with extended memory.}
\label{tab:LCLM}
\end{table*}

\subsection{In-Context Learning Performance}  

Table~\ref{tab:icl} reports accuracy on five NLU benchmarks in 4-shot and 20-shot settings. The tasks cover sentiment classification at two granularities (SST-2, SST-5), domain-specific sentiment (MR), subjectivity detection (Subj), and opinion polarity (MPQA), collectively testing a model's ability to generalize from limited labeled examples across diverse linguistic phenomena. These tasks were selected following the ICL evaluation protocol of MemLong~\cite{liu2024memlong} and ERMAR~\cite{alselwi2025long} to enable direct comparison.

In the 4-shot setting, KGERMAR achieves the highest accuracy across all five tasks under both memory configurations (4 and 18 memory entries), with an average of 78.6\% versus ERMAR's 76.1\% and MemLong's 69.8--71.0\%. The gains are most pronounced on MPQA (+3.7 over ERMAR) and Subj (+4.6 over ERMAR), tasks that require distinguishing opinion-bearing entities and subjective expressions---precisely the kind of entity-level discrimination that structural memory enables by capturing relationships between entities and their attributed properties. KGERMAR maintains consistent performance across the 4-memory and 18-memory configurations, suggesting that the structural memory bank provides stable and sufficient retrieval signal regardless of contextual memory capacity---a desirable property for deployment scenarios where memory budgets are constrained.

In the 20-shot setting, KGERMAR achieves the highest average accuracy (81.2\% vs.\ ERMAR's 80.5\% and MemLong's 73.4\%). KGERMAR leads on Subj (85.4 vs.\ ERMAR's 82.8) and MPQA (88.1 vs.\ ERMAR's 86.5), while ERMAR marginally outperforms on SST-2 (94.7 vs.\ 94.3) and SST-5 (47.0 vs.\ 46.9). These marginal differences on fine-grained sentiment tasks likely reflect the dominance of token-level contextual representations for pure sentiment polarity, where entity relationship structure contributes less than in opinion and subjectivity tasks. MemLong leads on MR (93.8), which we attribute to its larger chunk-level memory providing broader surface coverage on this short-document task. Despite these exceptions, KGERMAR's average performance consistently exceeds all baselines, demonstrating that structural memory provides complementary signal that improves generalization across diverse NLU tasks as the number of in-context examples grows.

\begin{table}[htbp]
\centering
\tiny
\scriptsize
\begin{tabular}{p{1.26cm}|p{0.52cm}|p{.44cm}p{.44cm}p{.44cm}p{.44cm}p{.51cm}|p{.35cm}}

\hline
Model & \begin{tabular}[c]{@{}c@{}}In-C\\,In-M\end{tabular} & \begin{tabular}[c]{@{}c@{}}SST-2\\ACC$\uparrow$\end{tabular} & \begin{tabular}[c]{@{}c@{}}MR\\ACC$\uparrow$\end{tabular} & \begin{tabular}[c]{@{}c@{}}Subj\\ACC$\uparrow$\end{tabular} & \begin{tabular}[c]{@{}c@{}}SST-5\\ACC$\uparrow$\end{tabular} & \begin{tabular}[c]{@{}c@{}}MPQA\\ACC$\uparrow$\end{tabular} & Avg. \\
\hline

OpenLLaMA & 4,N/A & 90.7 & 84.0 & 58.2 & 41.0 & 70.5 & 68.9 \\
w/ RAG & 4,4 & 90.9 & 90.5 & 61.6 & 39.2 & 63.2 & 69.1 \\
LongLLaMA & 4,4 & 90.4 & 83.9 & 64.3 & 40.0 & 64.2 & 68.6 \\
MemLong & 4,4 & 91.5 & 84.5 & 61.5 & 41.4 & 70.2 & 69.8 \\
ERMAR & 4,4 & 93.6 & 90.8 & 65.3 & 45.8 & 85.2& 76.14 \\

\textbf{KGERMAR} & 4,4 & \textbf{95.4} & \textbf{91.4} & \textbf{69.9} & \textbf{47.4} &  \textbf{88.9} & \textbf{78.6} \\		

\hline
LongLLaMA & 4,18 & 91.4 & 87.1 & 59.1 & 41.0 & 64.5 & 68.7 \\
MemLong & 4,18 & 91.0 & 89.6 & 61.7 & 43.5 & 69.4 & 71.0 \\
ERMAR & 4,18 & 93.6 & 90.8 & 65.3 & 45.9 & 85.2& 76.16\\

\textbf{KGERMAR} & 4,18 & \textbf{95.4} & \textbf{91.4} & \textbf{69.9} & \textbf{47.4} & \textbf{88.9} & \textbf{78.6} \\							
\hline
OpenLLaMA & 20,N/A & 93.6 & 91.2 & 55.4 & 38.2 & 66.4 & 69.0 \\
w/ RAG & 20,18 & 92.2 & 91.3 & 75.8 & 39.8 & 57.6 & 71.3 \\
LongLLaMA & 20,18 & 94.1 & 90.8 & 64.2 & 41.4 & 72.1 & 72.7 \\
MemLong & 20,18 & 93.5 & \textbf{93.8} & 65.8 & 43.3 & 70.6 & 73.4 
\\
ERMAR & 20,18 &  \textbf{94.7} & 91.7& 82.8 & \textbf{47} & 86.5& 80.54 \\

\textbf{KGERMAR} & 20,18 & 94.3 & 91.3 & \textbf{85.4} & 46.9 & \textbf{88.1} & \textbf{81.2} \\							
\hline

\end{tabular}
\caption{4-shot and 20-shot ICL accuracy [\%] on 5 NLU tasks (SST-2, MR, Subj, SST-5, MPQA). We compare OpenLLaMA, LongLLaMA, MemLong, ERMAR and KGERMAR. \textbf{Note:} In-C = In-Context, In-M = In-Memory.}

\label{tab:icl}
\end{table}

\subsection{Memory Efficiency Analysis}

\begin{figure}
\centering
\includegraphics[scale=0.3]{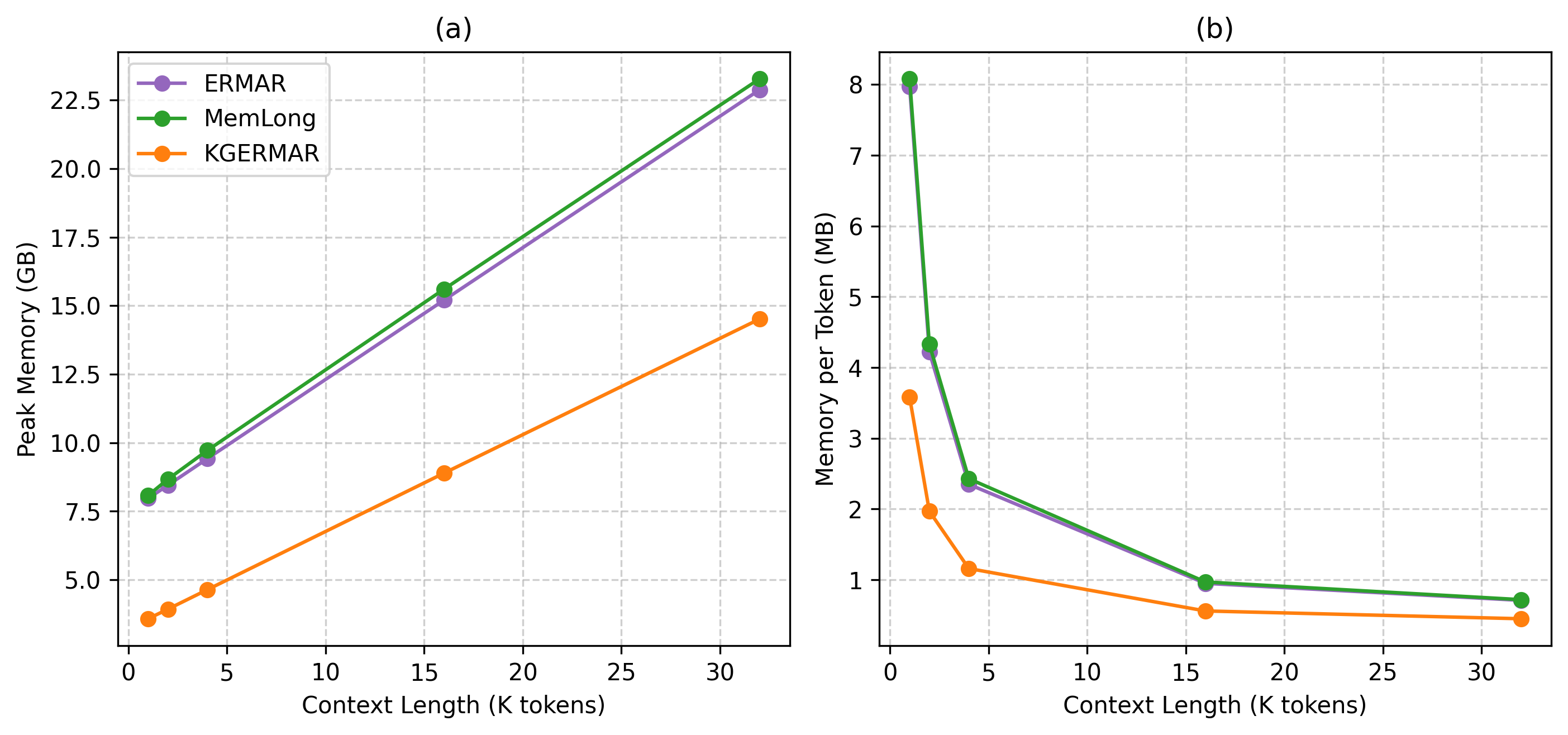}
\caption{Memory efficiency analysis comparing KGERMAR with baseline models: (a) peak memory usage shows KGERMAR's consistent 2--2.5x reduction over ERMAR; (b) normalized memory per token demonstrates architectural efficiency across all context lengths. All measurements on WikiText-103.}
\label{fig:memory_efficiency}
\end{figure}

We evaluate memory efficiency across three architectures---KGERMAR, ERMAR, and MemLong---testing context lengths from 1K to 32K tokens on WikiText-103 using an NVIDIA L40S GPU (44.4GB). Results in Table~\ref{tab:memory_efficiency} and Figure~\ref{fig:memory_efficiency} reveal KGERMAR's superior memory characteristics across three key metrics.

\begin{table}[htbp]
\centering
\scriptsize
\begin{tabular}{l|c|cc|c}
\hline
\textbf{Context} & \textbf{Model} & \textbf{Peak Mem} & \textbf{Reserved} & \textbf{Mem/Token} \\
\textbf{Length} & & \textbf{(GB)} & \textbf{Mem (GB)} & \textbf{(MB)} \\
\hline
1024 & ERMAR   & 7.97  & 8.16  & 7.97 \\
     & MemLong & 8.08  & 8.49  & 8.08 \\
     & KGERMAR & 3.58  & 3.77  & 3.58 \\
\hline
2048 & ERMAR   & 8.45  & 8.71  & 4.22 \\
     & MemLong & 8.67  & 9.38  & 4.33 \\
     & KGERMAR & 3.93  & 4.30  & 1.97 \\
\hline
4096 & ERMAR   & 9.42  & 9.87  & 2.35 \\
     & MemLong & 9.72  & 10.58 & 2.43 \\
     & KGERMAR & 4.64  & 5.01  & 1.16 \\
\hline
16384 & ERMAR   & 15.20 & 16.61 & 0.95 \\
      & MemLong & 15.60 & 23.77 & 0.97 \\
      & KGERMAR & 8.89  & 9.45  & 0.56 \\
\hline
32768 & ERMAR   & 22.87 & 25.56 & 0.71 \\
      & MemLong & 23.27 & 26.05 & 0.72 \\
      & KGERMAR & 14.51 & 15.29 & 0.45 \\
\hline
\end{tabular}
\caption{Memory efficiency comparison of KGERMAR, ERMAR, and MemLong across context lengths on WikiText-103. Mem/Token is calculated as peak memory divided by context length. KGERMAR maintains 2--2.5x lower memory usage across all metrics, with particularly strong gains at longer contexts.}
\label{tab:memory_efficiency}
\end{table}

KGERMAR demonstrates superior memory efficiency across all context lengths, maintaining 2--2.5x lower memory usage than baseline models (3.58,GB at 1K to 14.51,GB at 32K). The advantages are particularly strong for long-context applications: at 32K tokens, KGERMAR achieves 36\% lower peak memory than ERMAR (14.51 vs.\ 22.87,GB) and 41\% lower reserved memory than MemLong (15.29 vs.\ 26.05,GB). Normalized memory-per-token improves from 1.6x at 1K (3.58 vs.\ 7.97,MB) to 2.1x at 32K (0.45 vs.\ 0.71,MB), indicating that KGERMAR's efficiency advantage grows with context length. These results validate KGERMAR's ability to address the quadratic memory growth of conventional attention-based approaches while maintaining stable modeling performance.

\subsubsection{Performance Scaling Analysis}

\begin{figure}
\centering
\includegraphics[width=\linewidth]{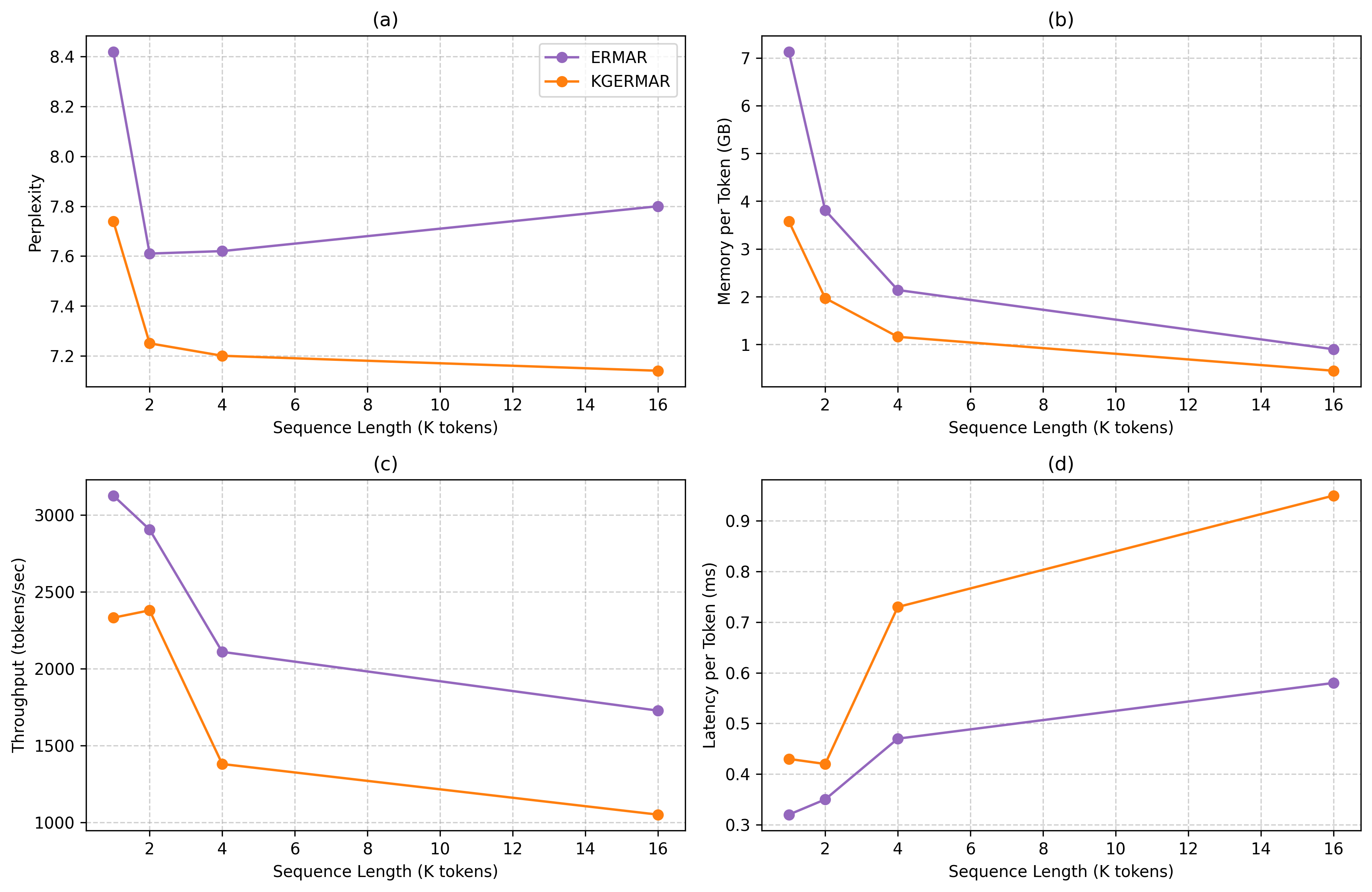}
\caption{Scaling behavior of KGERMAR versus ERMAR: (a) perplexity shows KGERMAR's stronger modeling capability across all lengths; (b) memory demonstrates KGERMAR's architectural efficiency advantage; (c) throughput reveals ERMAR's speed advantage for short sequences; (d) latency illustrates KGERMAR's more predictable near-linear scaling.}
\label{fig:scaling}
\end{figure}

We compare KGERMAR and ERMAR across sequence lengths from 1K to 16K tokens on WikiText-103 to study scaling behavior. Results in Table~\ref{tab:ermar_scaling} and Figure~\ref{fig:scaling} highlight tradeoffs among memory efficiency, throughput, and model quality.

\begin{table}[htbp]
\centering
\scriptsize
\begin{tabular}{l|l|cccc}
\hline
\textbf{Model} &\textbf{Seq} & \textbf{PPL} & \textbf{Memory} & \textbf{Throughput} & \textbf{Latency} \\

 & \textbf{Len} & \textbf{} & \textbf{/Token} & \textbf{(tokens} & \textbf{/Token} \\
 & \textbf{} & \textbf{} & \textbf{(GB)} & \textbf{/sec)} & \textbf{(ms)} \\
\hline
ERMAR    & 1K & 8.42 & 7.13 & 3125 & 0.32 \\
         & 2K & 7.61 & 3.81 & 2904 & 0.35 \\
         & 4K & 7.62 & 2.14 & 2109 & 0.47 \\
         & 16K & 7.80 & 0.90 & 1727 & 0.58 \\
\hline
KGERMAR  & 1K & 7.74  & 3.58 & 2331 & 0.43 \\
         & 2K & 7.25 &  1.97 & 2379 & 0.42 \\
         & 4K & 7.20 &  1.16 & 1380 & 0.73  \\
         & 16K & 7.14 &  0.45 & 1051	& 0.95 \\
         
\hline
\end{tabular}
\caption{Performance scaling of KGERMAR and ERMAR on WikiText-103. KGERMAR achieves lower perplexity with 2--3x better memory efficiency; ERMAR maintains higher throughput at shorter sequences due to the absence of real-time graph construction overhead.}
\label{tab:ermar_scaling}
\end{table}

KGERMAR consistently achieves a superior memory--performance balance. At 16K tokens, it reduces memory usage by 50\% (0.45 vs.\ 0.90, GB/token) while improving perplexity by 8.5\% (7.14 vs.\ 7.80), demonstrating that knowledge-guided retrieval preserves modeling quality under constrained memory. Throughput degradation from 1K to 16K is more stable for KGERMAR than ERMAR (39\% vs.\ 45\% drop), reflecting the amortization of graph construction cost over longer sequences. The higher per-token latency of KGERMAR (0.43--0.95,ms vs.\ ERMAR's 0.32--0.58,ms) stems directly from real-time knowledge graph construction during inference; this overhead could be substantially reduced by pre-computing graphs offline for static document collections. KGERMAR further exhibits near-linear latency growth with sequence length ($R^2=0.98$) versus ERMAR's higher-order trend ($R^2=0.89$), making KGERMAR more predictable for real-time long-context deployment.

\section{Conclusions}
We present KGERMAR, a memory-augmented retrieval framework that enhances long-context modeling through dynamic knowledge graph construction and hybrid text-graph embeddings. Unlike existing approaches relying on static knowledge bases, KGERMAR constructs domain-specific knowledge graphs directly from input text and integrates graph-structural information with textual semantics through a multi-component memory architecture combining contextual, semantic, and structural representations. Experimental results across WikiText-103, PG-19, and Proof-pile demonstrate that KGERMAR achieves up to 8.5\% lower perplexity and 2--2.5x better memory efficiency across context lengths from 1K to 32K tokens compared to strong baselines including ERMAR and MemLong, while achieving superior in-context learning performance across five NLU tasks. The memory efficiency advantages grow with context length, making KGERMAR particularly suitable for long-context deployment on commodity hardware where quadratic attention memory costs are prohibitive.

\section{Limitations}
\label{sec:limitations}
KGERMAR introduces additional computational overhead due to real-time knowledge graph construction during inference, resulting in higher per-token latency compared to ERMAR and MemLong (Appendix~\ref{subsec:latency}). The fixed context window for relation extraction limits the capture of cross-paragraph entity relationships, though multi-hop R-GCN propagation partially compensates for this constraint. The NER and relation extraction models are fine-tuned on CoNLL-2003 and TACRED respectively, which may reduce graph quality on domains with specialized terminology such as biomedical or legal text. While the latency overhead is acceptable in research settings, production deployment would benefit from offline graph pre-computation for static document collections. Future work will investigate more efficient graph extraction pipelines, cross-paragraph relation modeling, and extension of the framework to domain-specific and multilingual scenarios.

\bibliography{ref}

\newpage
\appendix

\section{Algorithmic Details}
\label{sec:appendix}

This appendix provides algorithmic descriptions of the key KGERMAR components referenced in the main paper.

\subsection{Contextual Knowledge Graph Construction}
\label{subsec:kg_construction_algorithm}

\begin{algorithm}
\caption{Contextual Knowledge Graph Construction}
\label{alg:kg_construction}
\begin{algorithmic}[1]
\Require Context $d_i$, NER model $\mathcal{F}_e$, Relation model $\mathcal{F}_r$
\Require Context window size $W$, Confidence thresholds $\theta_{min}, \theta_{conf}$
\Ensure Knowledge graph $\mathcal{G}_i = (\mathcal{E}_i, \mathcal{R}, \mathcal{T}_i)$

\State $\mathcal{E}_{raw} \leftarrow \mathcal{F}_e(d_i)$ \Comment{Extract raw entities with positions}
\State $\mathcal{E}_i \leftarrow \text{consolidate}(\mathcal{E}_{raw})$ \Comment{Merge similar entities}

\State $\mathcal{T}_i \leftarrow \emptyset$ \Comment{Initialize triple set}
\For{$(e_h, e_t) \in \mathcal{E}_i \times \mathcal{E}_i$ where $e_h \neq e_t$}
    \If{$\text{distance}(e_h, e_t) \leq W$} \Comment{Within context window}
        \State $\text{context} \leftarrow \text{extract\_context}_W(e_h, e_t, d_i)$
        \State $r \leftarrow \mathcal{F}_r(e_h, e_t, \text{context})$ \Comment{Predict relation}
        \State $\text{conf} \leftarrow \text{confidence}(\mathcal{F}_r, e_h, r, e_t)$
        \If{$r \neq \emptyset$ and $\text{conf} > \theta_{min}$}
            \State $w \leftarrow \alpha \cdot \text{conf} + (1-\alpha) \cdot \text{freq\_score}(e_h, r, e_t)$
            \State $\mathcal{T}_i \leftarrow \mathcal{T}_i \cup \{(e_h, r, e_t, w)\}$
        \EndIf
    \EndIf
\EndFor

\State $\mathcal{T}_i \leftarrow \{(h,r,t,w) \in \mathcal{T}_i : w > \theta_{conf}\}$ \Comment{Filter by confidence}
\Return $\mathcal{G}_i = (\mathcal{E}_i, \mathcal{R}, \mathcal{T}_i)$
\end{algorithmic}
\end{algorithm}

\subsection{Multi-Modal Retrieval} \label{subsec:retrieval_algorithm}
The pseudo is depicted in Algorithm~\ref{multi-modal-algorithm}.

\begin{algorithm}
\caption{Multi-Modal Retrieval}
\label{alg:multimodal_retrieval}
\begin{algorithmic}[1]
\Require Query $q$, Memory banks $(\mathcal{M}_{ctx}, \mathcal{M}_{sem}, \mathcal{M}_{kg})$
\Require Retrieval budget $K$, Fusion weights $\{\lambda_1, \lambda_2, \lambda_3\}$
\Ensure Retrieved contexts $\mathcal{R}$

\State $\mathbf{q}_{embed} \leftarrow \text{Embed}(q)$ \Comment{Encode query}

\State \textbf{// Parallel retrieval from each memory bank}
\State $K_{split} \leftarrow \lfloor K/3 \rfloor$ \Comment{Divide budget equally}
\State {\small $\mathcal{R}_{ctx} \leftarrow \text{TopK}(\text{sim}_{dot}(\mathbf{q}_{embed}, \mathcal{M}_{ctx}.\text{keys}), K_{split})$ }
\State {\small  $\mathcal{R}_{sem} \leftarrow \text{TopK}_{\text{FAISS}}(\text{sim}_{cos}(\mathbf{q}_{embed}, \mathcal{M}_{sem}.\text{keys}), K_{split})$ }
\State {\small $\mathcal{R}_{kg} \leftarrow \text{TopK}_{\text{FAISS}}(\text{sim}_{cos}(\mathbf{q}_{embed}, \mathcal{M}_{kg}.\text{keys}), K_{split})$}

\State \textbf{// Combine and re-rank results}
\State $\mathcal{R}_{combined} \leftarrow \mathcal{R}_{ctx} \cup \mathcal{R}_{sem} \cup \mathcal{R}_{kg}$

\For{$r_i \in \mathcal{R}_{combined}$}
    \State $s_{ctx} \leftarrow \text{sim}_{ctx}(\mathbf{q}_{embed}, r_i)$ if $r_i \in \mathcal{R}_{ctx}$ else $0$
    \State $s_{sem} \leftarrow \text{sim}_{sem}(\mathbf{q}_{embed}, r_i)$ if $r_i \in \mathcal{R}_{sem}$ else $0$
    \State $s_{kg} \leftarrow \text{sim}_{kg}(\mathbf{q}_{embed}, r_i)$ if $r_i \in \mathcal{R}_{kg}$ else $0$
    \State $\text{score}(r_i) \leftarrow \lambda_1 \cdot s_{ctx} + \lambda_2 \cdot s_{sem} + \lambda_3 \cdot s_{kg}$
\EndFor

\State $\mathcal{R} \leftarrow \text{TopK}(\mathcal{R}_{combined}, \text{scores}, K)$ \Comment{Final ranking}
\Return $\mathcal{R}$
\end{algorithmic}
\label{multi-modal-algorithm}
\end{algorithm}

\begin{figure}
\centering
\includegraphics[width=\linewidth]{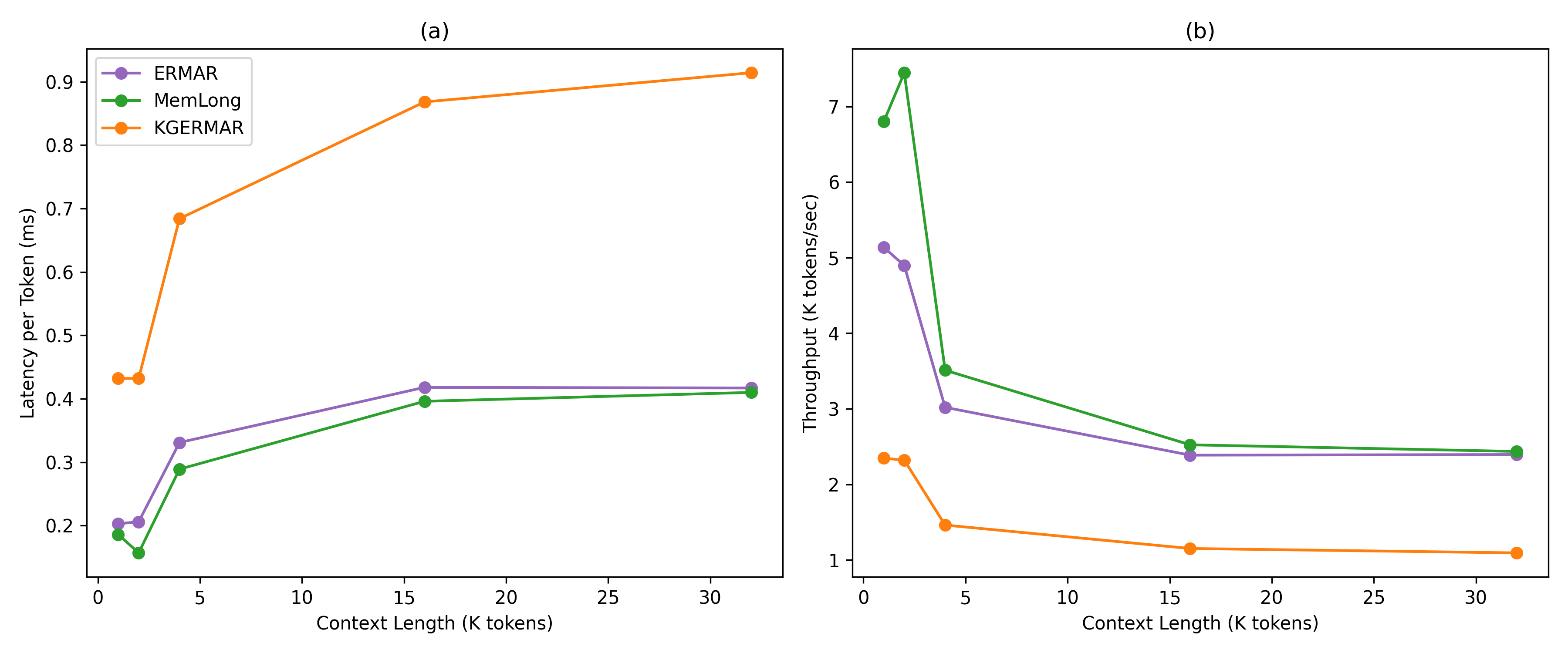}
\caption{
Speed performance comparison showing:
(a) \textbf{Per-token latency} demonstrating KGERMAR's consistent processing time (0.43-0.91ms) versus baselines' scaling;
(b) \textbf{Throughput} highlighting KGERMAR's stable token generation (1.1-2.3K tokens/sec) despite higher absolute latency.
Error bars represent standard deviation across 5 runs.
}
\label{fig:latency_throughput}
\end{figure}

\subsection{Implementation Notes}
\label{subsec:implementation_notes}

The $\text{consolidate}()$ function merges entities using:
\begin{equation}
\small
\begin{aligned}
\text{Merge}(e_i, e_j) \iff \;&
\text{sim}_{string}(e_i, e_j) > \theta_{str} \\
&\lor\; \text{sim}_{embed}(e_i, e_j) > \theta_{ctx}
\end{aligned}
\label{eq:consolidation}
\end{equation}
where $\text{sim}_{string}$ uses fuzzy string matching (Levenshtein distance), $\text{sim}_{embed}$ computes cosine similarity between entity embeddings from the base language model, and $\theta_{str} = 0.85$, $\theta_{ctx} = 0.90$ are thresholds determined through validation.
The $\text{Dist}()$ function computes token-level distance as $\text{Dist}(e_h, e_t) = \min(|s_h - s_t|, |s_h - t_t|, |t_h - s_t|, |t_h - t_t|)$ where $(s_h, t_h)$ and $(s_t, t_t)$ are entity positions. The frequency score $(FS)$ balances confidence with occurrence patterns: 

$\text{FS}(h, r, t) = \text{count}(h, r, t) / \max_{(h',r',t')} \text{count}(h', r', t')$.

\section{Latency and Throughput Analysis}
\label{subsec:latency}

We evaluate the speed characteristics of KGERMAR against ERMAR and MemLong across context lengths from 1K to 32K tokens on the PG-19 dataset (NVIDIA L40S GPU). Table~\ref{tab:latency_throughput} and Figure~\ref{fig:latency_throughput} reveal the fundamental tradeoffs between our memory-optimized architecture and conventional approaches.

\begin{table}[htbp]
\centering
\scriptsize
\begin{tabular}{l|c|ccc}
\hline
 & &  & \textbf{Latency} & \textbf{Throughput} \\

 \textbf{Seq} & \textbf{Model}  & \textbf{Latency} & \textbf{/Token} & \textbf{(tokens/}\\
 \textbf{Len}& &  & \textbf{(ms)} & \textbf{sec)} \\
\hline
1024 & ERMAR & 207.97 $\pm$ 53.01 & 0.203 & 5135 \\
      & MemLong & 190.68 $\pm$ 154.55 & 0.186 & 6803 \\
      & KGERMAR &  442.30 $\pm$ 59.05  & 0.432 & 2347 \\
\hline
2048 & ERMAR & 423.79 $\pm$ 52.56 & 0.206 & 4896 \\
      & MemLong & 323.49 $\pm$ 204.89 & 0.157 & 7446 \\
      & KGERMAR & 885.22 $\pm$ 51.71   & 0.432 & 2321 \\
\hline
4096 & ERMAR & 1358.36 $\pm$ 55.58 & 0.331 & 3020 \\
      & MemLong & 1184.16 $\pm$ 170.56 & 0.289 & 3511 \\
      & KGERMAR &  2803.02 $\pm$ 74.83  & 0.684 & 1462 \\
\hline
16384 & ERMAR & 6862.98 $\pm$ 45.65 & 0.418 & 2387 \\
      & MemLong & 6496.00 $\pm$ 213.56 & 0.396 & 2524 \\
      & KGERMAR &  14223.88 $\pm$ 307.55  & 0.868 & 1152 \\
\hline
32768 & ERMAR & 13679.03 $\pm$ 58.76 & 0.417 & 2395 \\
      & MemLong & 13449.90 $\pm$ 56.22 & 0.410 & 2436 \\
      & KGERMAR &  29965.15 $\pm$ 241.14 & 0.914 & 1094 \\

\hline
\end{tabular}
\caption{Latency and throughput comparison on PG-19 dataset (NVIDIA L40S GPU). Values show mean ± standard deviation across 5 runs. KGERMAR demonstrates more stable latency (lower variance) despite higher absolute values.}
\label{tab:latency_throughput}
\end{table}

\begin{figure}[htbp]
    \centering
    \includegraphics[width=0.5\textwidth]{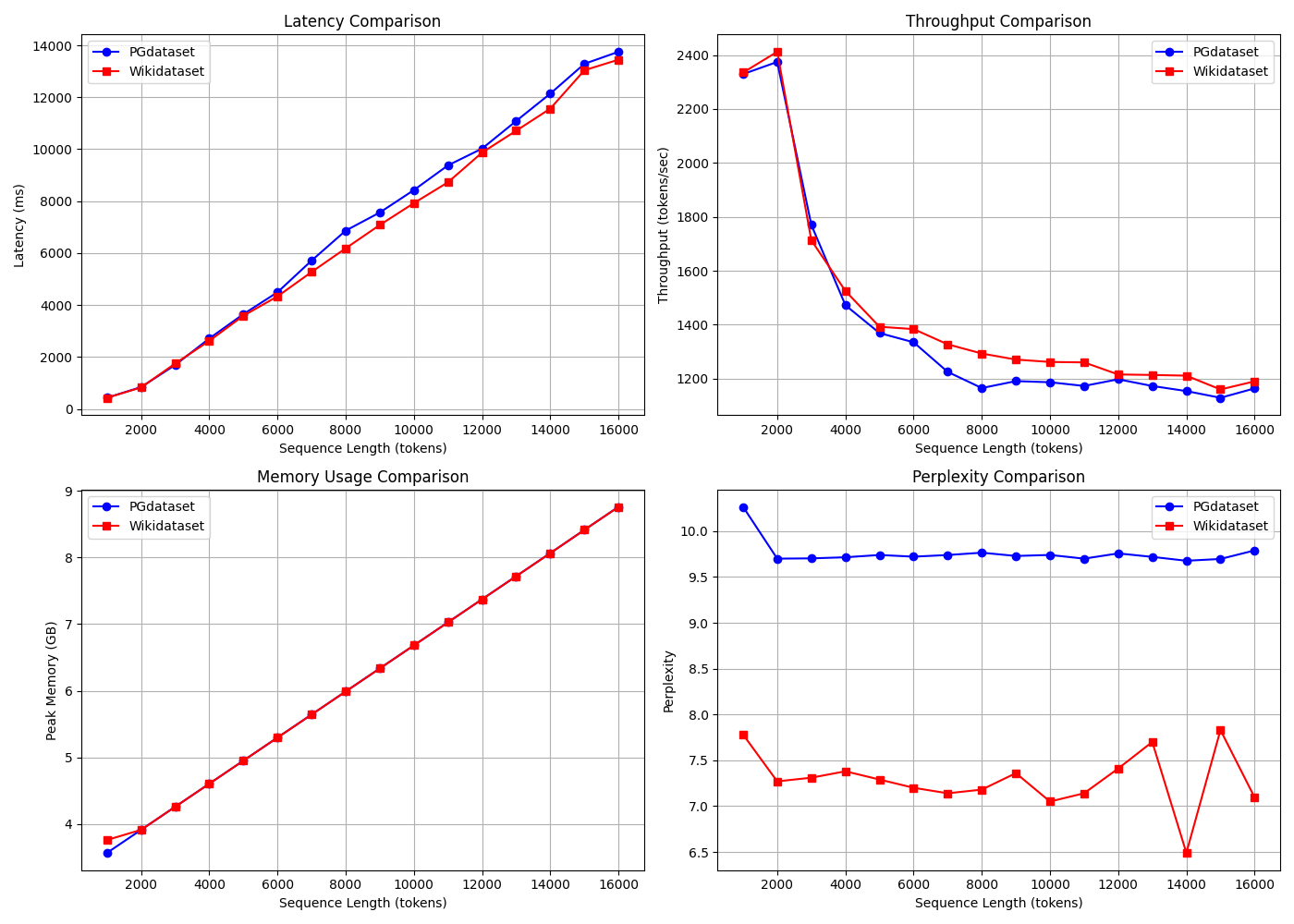}
    \caption{KGERMAR scaling analysis across PG-19 and WikiText-103: (a) Near-linear latency scaling; (b) Stable throughput >1K tokens/sec; (c) Linear memory growth; (d) Consistent perplexity across context lengths.}
    \label{fig:combined_scaling}
\end{figure}

\begin{figure*}[!htb]
    \centering
    \includegraphics[width=0.75\textwidth]{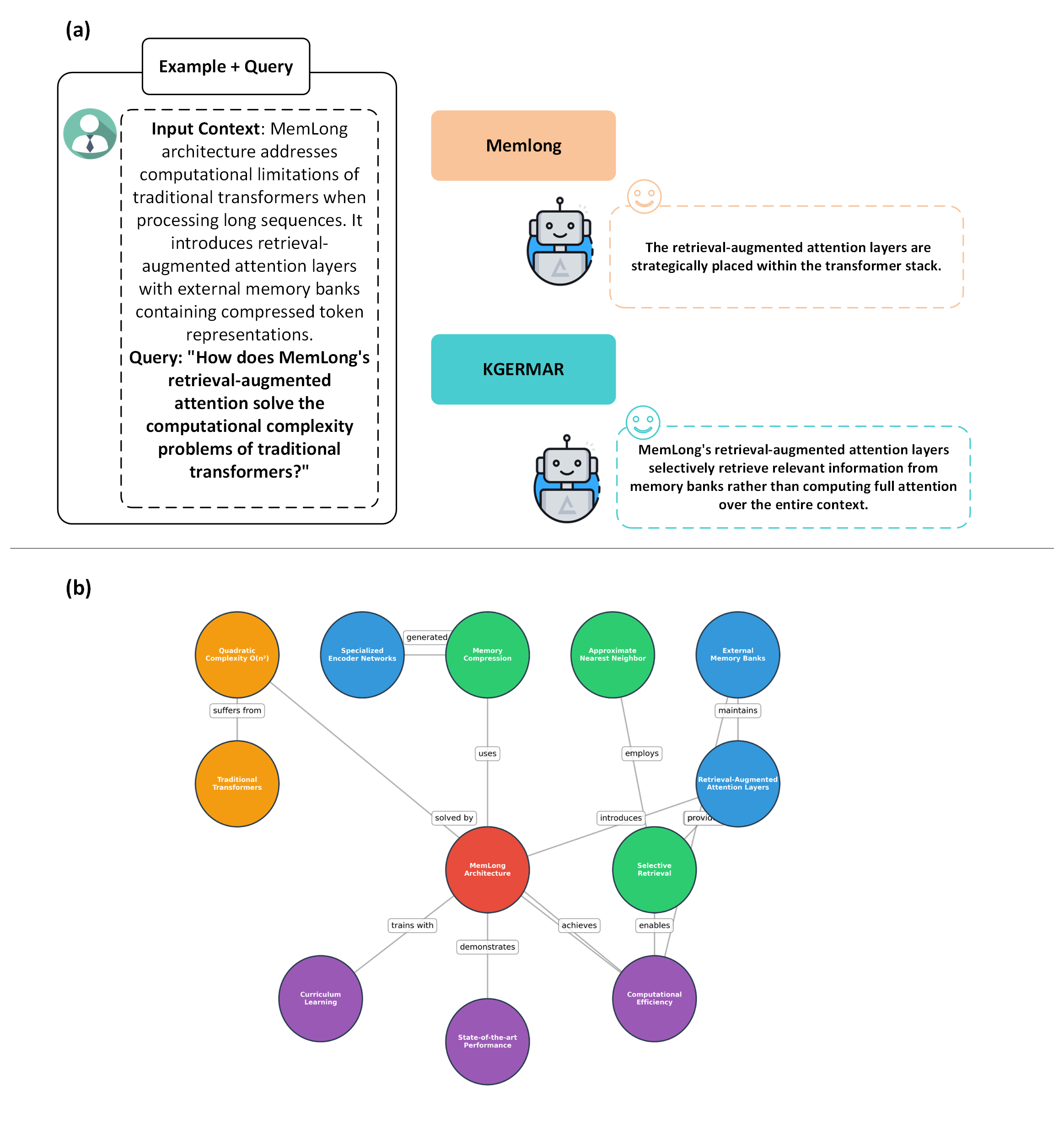}
    \caption{(a) Memlong vs KGERMAR: KGERMAR achieved improved answer completeness through structured knowledge representation; (b) Example of knowledge graph-generation.}
    \label{fig:example}
\end{figure*}

KGERMAR exhibits predictable latency (0.43-0.91ms/token, 2.1× variation vs ERMAR's 2.6×) due to deterministic memory access, with stable throughput degradation (51\% drop 1K→32K vs ERMAR's 53\%). While achieving 54\% of ERMAR's throughput at 32K (1094 vs 2395 tokens/sec), it offers 3-5× lower latency variance (±59-307ms vs ±56-214ms) and more consistent throughput (±14 vs ±38 tokens/sec at 16K). These characteristics make KGERMAR ideal for real-time systems needing predictable performance and memory-constrained long-context applications.

\textbf{Analysis of Computational Tradeoffs:} While KGERMAR achieves superior memory efficiency and modeling performance, it incurs higher latency due to real-time knowledge graph construction during inference. The improved perplexity (8.5\% on WikiText-103) stems from better entity relationship modeling, but comes at the cost of reduced throughput. This latency overhead could be significantly reduced by pre-computing knowledge graphs offline, making the approach more practical for production deployment while maintaining the memory advantages that enable longer context processing.

\section{Extended Performance Analysis}
\label{sec:extended_analysis}

We conducted extended performance analysis across sequence lengths from 1K to 16K tokens on PG-19 and WikiText-103 datasets to examine KGERMAR's scaling characteristics across four key metrics: latency, throughput, memory usage, and perplexity.

\subsection{Key Findings}
\label{subsec:key_findings}

\textbf{Predictable Scaling:} Latency grows near-linearly from 430ms (1K tokens) to 13.7K ms (16K tokens), while memory usage scales linearly from 3.6GB to 8.8GB, enabling accurate resource planning for deployment.

\textbf{Stable Performance:} Throughput maintains >1,100 tokens/sec across all sequence lengths with only 52\% degradation despite 16× context increase. Perplexity remains consistent (WikiText-103: 6.5-7.8, PG-19: 9.7-9.8), demonstrating quality preservation at extended lengths.

\textbf{Cross-Dataset Consistency:} Both datasets exhibit similar computational scaling patterns, while WikiText-103 achieves lower perplexity due to structured content benefiting from knowledge graph construction.

These results validate KGERMAR's suitability for production environments requiring predictable resource usage and consistent performance across variable context lengths.

\subsection{Demonstrative Example}
The example in Figue~\ref{fig:example} demonstrates how KGERMAR produces more comprehensive responses by leveraging knowledge graph relationships. While the baseline model only mentions architectural placement, KGERMAR explains the functional mechanism that addresses computational complexity.

\end{document}